\documentstyle[12pt,aps,psfig]{revtex}
\newcommand{\be}{\begin{equation}}
\newcommand{\ee}{\end{equation}}   
\newcommand{\J}{{\rm J}/\Psi}

\newcommand{\bea}{\begin{eqnarray}}
\newcommand{\eea}{\end{eqnarray}}

\def\Journal#1#2#3#4{{#1} {\bf #2}, #3 (#4)}

\def\NP{{\em Nucl. Phys.}}
\def\PL{{\em Phys. Lett.} B}

\def\PRL{\em Phys. Rev. Lett.}
\def\PRD{{\em Phys. Rev.} D}

\def\ZPC{{\em Z. Phys.} C}

\def\MPLA{{\em Mod. Phys. Lett. A}}

\begin{document}
\title{J/$\Psi$ production, $\chi$ polarization
and Color Fluctuations}

\author{L. Gerland${}^a$, L. Frankfurt${}^{a,b}$, M. Strikman${}^c$,
\\H. St\"ocker${}^a$, W. Greiner${}^a$}
\address{{\it a} Institut f\"ur Theoretische Physik der 
J.W.Goethe-Universit\"at\\
Robert-Mayer-Str. 8-10, D-60054 Frankfurt, Germany
\\
{\it b} School of Physics and Astronomy, Tel Aviv University,
69978 Ramat Aviv, Tel Aviv, Israel
\\
{\it c} Department of Physics, Pennsylvania State University\\
University Park, PA 16802, USA}

\date{\today}

\maketitle
\begin{abstract}
The hard contributions to
the heavy quarkonium-nucleon cross sections
are calculated based on the QCD factorization theorem
and the nonrelativistic quarkonium model.
We evaluate the nonperturbative
part of these cross sections which
dominates at $\sqrt{s_{NN}}\approx 20$ GeV 
at the Cern Super Proton Synchrotron (SPS)  
and becomes a correction  at $\sqrt{s_{NN}}\approx 6$ TeV 
at the CERN Large Hadron Collider (LHC). 
$\J$ {\it production} at the
CERN SPS is well described by hard QCD,
when the larger absorption cross sections of the $\chi$ states  
predicted by QCD are taken into account.
We predict an $A$-dependent polarization of the $\chi$ states. 
The expansion of small wave packets is discussed.
\end{abstract}

\newpage

The aim of the present paper is to investigate color 
coherent phenomena in the propagation of hadronic states with hidden 
charm in $pA$ and $AB$ collisions. We restrict the
consideration to the central rapidity region of nuclear beams at 
$\sqrt{s}\approx 20$ AGeV (CERN SPS) where formation time effects in 
$\J$-production are corrections.
Qualitatively new effects for $\sqrt{s}\approx 200$ AGeV (RHIC) and 
$\sqrt{s}\approx 6$ ATeV (LHC) energies will be discussed also.
We assume that hadronic states with hidden flavor are produced predominantly
in hard parton collisions and calculate the absorption factor $S$ of the total
yield of heavy quarkonia which is not distorted by the initial state
interaction, cf.~\cite{bodwin}.  

Under these assumptions $S$ can be directly measured by the ratio 
of the yields of heavy quarkonium to the dimuon yield~\cite{carlos}. We want to 
emphasize that the agreement of our calculations of the J/$\Psi$ production 
in $pA$ and in $AB$ collisions with the data strongly supports the idea that 
hidden charm states are  perturbatively produced in $NN$ collisions.
 
The QCD factorization theorem is used to evaluate the PQCD cross 
sections of 
heavy quarkonium interactions with ordinary hadrons. However, the charmonium 
states (here denoted $X$) are not sufficiently small to ignore nonperturbative 
QCD physics.
Thus, we evaluate the nonperturbative QCD contribution to the cross sections 
of charmonium-nucleon interaction by using an interpolation between 
known cross sections. The $\J$-$N$ cross section evaluated in this paper is
in reasonable agreement with SLAC data \cite{slac}.

PQCD predictions for the cross section of  $\Upsilon$-$N$
scattering should be more reliable since (differenctly from the J/$\Psi$'s) 
the spatial size of $\Upsilon$'s is significantly smaller than the typical 
size for soft hadron interactions.

These $X$-$N$ cross sections are used to evaluate nuclear effects 
for the production of $X$ states. We find that the suppression of the $\J$ in 
$pA$ and $AB$ collisions is reasonably well described within the present 
approach.

We predict a new QCD effect -- the polarization of produced $\chi_c$ states:
$c\bar c$-states with nonzero orbital momentum will be polarized due to the 
longitudinally directed nuclear (color filter) target. Their polarization
is evaluated quantitatively. Random interactions with comovers 
will depolarize the $\chi$ states. Thus, measuring 
the $\chi$ polarization may turn out to be an effective method to
investigate their interactions with comovers. 

The production of charmonium states in $AB$ collisions has been
proposed (via $\J$ suppression) as signal for Debye screening in deconfined
matter~\cite{matsui}. However, secondary collisions of $\J$'s with hadrons
have been suggested early on as a competing origin of $\J$ 
suppression~\cite{neubauer,gavin}. 
This topic has been considered also in~\cite{kharzeev,kopelio}, basing on   
QCD motivated models (but not on the QCD factorization theorem as 
in our paper). In~\cite{kharzeev} (but not in~\cite{kopelio}) it
is also concluded that PQCD predicts a too small J/$\Psi$-$N$ cross 
section to explain the observed absorption of $\J$ at CERN
energies. In contrast to the "preresonance" hypothesis suggested 
in~\cite{kharzeev} and to the model in~\cite{kopelio} we 
explain the $\J$ suppression observed in $AB$ collisions as a consequence 
of the predicted large absorption of $\chi$ states.
We found that at RHIC energies 
perturbative and nonperturbative QCD contributions to the $\J$-$N$
cross section become comparable in variance to~\cite{kharzeev}. 
At LHC energies the hard contribution
will dominate (we ignore in this paper the unitarity corrections which 
will be important at LHC, cf.~\cite{sigma,felix}). The $\Upsilon$-$N$ 
interaction cross section is calculated for the first time in this paper.

The first evaluations of $\sigma_{tot}(\J$-$N)$
have been obtained by applying the Vector Dominance Model to $\J$ 
photoproduction data. This leads to $\sigma_{\J N} \sim 1$ mb
for $E_{inc} \sim 20$ GeV. However, the application of the VDM leads 
to a paradox~\cite{FS85} -- one obtains
$\sigma_{tot}(\Psi'$-$N) \approx 0.7\cdot \sigma_{tot}(\J$-$N)$, although, 
on the other hand, $r_{\Psi'} \approx 2 r_{\J}$. This clearly indicates 
that the charmonium states produced in photoproduction are in 
a smaller -- than average -- configuration. Therefore, the VDM 
significantly underestimates $\sigma_{tot}(\J$-$N)$~\cite{FS85}. 

Indeed, the $A$-dependence of the $\J$ production studied at SLAC
at $E_{inc} \sim 20$ GeV exhibits a significant absorption effect \cite{slac}
leading to $\sigma_{abs}(\J$-$N)= 3.5 \pm 0.8$ mb.
It was demonstrated in \cite{farrar} that, in the kinematic region at SLAC, 
the color coherence effects are still small on the internucleon
scale for the formation of $\J$'s and lead  only to a small increase 
of the value of $\sigma_{abs}(\J$-$N)$. 
So, in contrast to the findings at higher energies, at intermediate 
energies this process measures the {\it genuine} $\J$-$N$ 
interaction cross section at energies of $\sim $ 15-20 GeV \cite{farrar}.

To evaluate the nonperturbative QCD contribution we use an 
interpolation formula for the dependence of the cross section
on the transverse size $b$ of a quark-gluon configuration~\footnote{
Note that the deviation of $\sigma(b)$ from the estimate of PQCD occurs
for $b\ge 0.3$ fm which is the distance close to the 
Chiral Phase Transition
scale. We thank J. Bjorken for emphasizing this point.}.
Three reference points are used to fix our parametrization of 
the cross sections (cf. Tab.~\ref{meanb}):
\begin{itemize}
\item
based on the observation that the $\phi$-$N$ cross section is  nearly
half of the $\rho$-$N$ cross section we impose 
$\sigma(b_0\cdot m_{\rho}/ m_{\phi})=$10-12 mb, where $b_0$ is 0.6 fm.
\item
$b_0$ is approximately the transverse size of a $\pi$, so we choose
$\sigma(b_0)=25$ mb, as known from $\pi$-$N$ collisions.
\item
for configurations with $b\geq 1$ fm we set $\sigma(1\, {\rm fm})=40$ mb, 
the value reached when the two constituent quarks split to form two 
open charm mesons.
\end{itemize}
Thus, the $X$-$N$ cross sections is calculated via:
\be
\sigma=\int \sigma(b)\cdot |\Psi (x,y,z)|^2 {\rm d}x\, {\rm d}y\, 
{\rm d}z \quad .
\label{integral}
\ee
where $\Psi (x,y,z)$ is the charmonium wave function. In our calculations
we use the wave functions from a non-relativistic charmonium model
with a Cornell confining potential, $V(r)=-\frac{\kappa}{r}+\frac{r}{a^2}$,
see~\cite{werner} and refs. therein.

The calculated cross sections of the charmonium states are shown in 
Tab.~\ref{meanb}. For the $\J$ we found $\sigma_{\J N}= 3.6$ mb. This
is in good agreement with the data~\cite{slac}, $\sigma=3.5\pm0.8$ mb, 
for the $\J$-photoproduction!

Note the factor of $\approx 2$ between the $<b^2>$ values of the $\chi_{10}$
versus the $\chi_{11}$, respectively: the transverse size of
a fast $\chi$ depends drastically on the polarization
of the $\chi$ states. Following the above discussion one therefore  
can expect significant fluctuations of the strength of the interaction
due to this geometrically directed filtering,
even in those kinematic regions where the color coherent phenomena
seem to be at first sight only a small correction.
The values of the $\chi$-$N$ cross sections are given for different
magnetic quantum numbers.

We calculate $\sigma(X$-$N)$ using the assumption that PQCD is 
applicable in this case. Here we will ignore the differences
between bare quarks of the QCD Lagrangian and the constituent quarks. Then the 
expression for $\sigma(X$-$N)$ follows unambigously from the QCD factorization 
theorem \cite{sigma}. The numerical calculation shown in Tab.~\ref{meanb}
yields at CERN energies $\sigma(X$-$N)$ values which are significantly 
smaller than in~\cite{slac}. 
The $\sigma_{hard}$ values~\footnote{The calculated $\sigma_{hard}$ reflects 
the effective cross section for the interaction of a $Q\bar Q$ pair  with
nucleons of the residual nucleus, which will transform into
the corresponding $Q\bar Q$ state only after it has traversed the nucleus.}
 in Tab.~\ref{meanb} are also calculated with 
eq.~\ref{integral}, but here we integrate only over the region 0.1 fm 
$<b<0.2$ fm, because $\alpha_s$ increases with $b$. 
Here $b$ is the {\it transverse distance} between the heavy quarks transverse to 
the momentum of the heavy quarkonium. Thus, the region $b>0.2$ fm
cannot be calculated within PQCD. We choose the lower limit for the integral,
$b=0.1$ fm, because the quarkonium states will have a finite size  already at 
the production point. For the calculation of the hard cross section the 
$\sigma(b)$ in eq.~\ref{integral} is given by the function ~\cite{sigma}
\be
\sigma(b)=\frac{\pi^2}{3}b^2\alpha_s(Q^2)\cdot xG(x,Q^2)\quad .
\ee

The Bjorken x needed for the calculation of $\sigma_{hard}(X-N)$ is calculated
by $x={Q^2 \over 2m_N\nu}$, where $Q^2={9\over b^2}$, $m_N$ is the nucleon mass
and $\nu$ is the energy of the heavy Quarkonium state $X$ in the rest frame of
nucleon. The calculation in Tab.~\ref{meanb} was done for a state $X$ produced
in midrapidity at SPS energy and in the target fragmentation region at RHIC and
LHC. One can see that the hard contribution to the cross section is just a
correction at SPS energies, but at RHIC energies both contributions
become comparable and at LHC it dominates (we neglect here that the DGLAP
equation (Dokshitser-Gribov-Lipatov-Altarelli-Parisi) should be probably  
violated~\cite{felix}).

We follow the analysis of~\cite{kharzeev} (and refs. therein) to evaluate 
the fraction of $\J$'s (in $pp$ collisions) that come from the decays of
the $\chi$ and $\Psi'$. No experimental
information is available on the contribution from the $\chi(1^{+-})$, 
nor are any contributions of the higher $c \bar c$ states (D-wave, radial
excitations) known. Thus, the fraction of the directly produced $\J$'s used 
in eq.~\ref{mix} is {\bf an upper limit} for
the actual fraction of directly produced $\J$'s. Furthermore, the ratio of
directly produced $\J$'s to the yield of $\J$'s due to $\Psi'$ decays is
$\J$ : $\Psi'$ = 0.92 : 0.08 in $pp$ collisions~\cite{kharzeev}. So, the
suppression factor $S$ of $\J$'s produced in the nuclear medium is 
calculated as: 
\be
S=0.6\cdot ( 0.92\cdot S^{\J}+0.08\cdot S^{\Psi'})+0.4\cdot S^{\chi}\quad .
\label{mix}
\ee 
Here $S^X$ are the respective suppression factors of the different 
pure charmonium states $X$ in nuclear
matter. The $S^X$ are for minimum bias $pA$ collisions within the
semiclassical approximation (cf.~\cite{hufner}): 
\bea
S_A^X&=&\frac{\sigma(pA\rightarrow X)}{A \cdot \sigma(pN\rightarrow X)} =
{1\over A} \int {\rm d}^2B\,{\rm d}z\, \rho(B,z)\cdot \exp \left(-
\int_z^{\infty}\sigma_{X} \rho(B,z'){\rm d}z'\right) \quad . 
\label{glaub}
\eea 
Here $\rho(B,z)$ is the local nuclear groundstate density (we used the
standard parametrization from~\cite{devries}).

The charmonium states are produced as small ($r_{init}\approx 0.2$ fm)
configurations predominantly in gluon-gluon-fusion, then they evolve to 
their full size. 
Therefore, if the formation length of the charmonium states, $l_f$, becomes 
larger than the average internucleon distance
($l_f>r_{NN}\approx 1.8$ fm), one has to take into account the evolution of 
the cross sections with the distance from the production point~\cite{farrar}.

The formation length of the $\J$ is given by the energy denominator
$l_f\approx \frac{2p}{m^2_{\Psi'}- m^2_{\J}}$, where $p$ is the momentum of 
the $\J$ in the rest frame of the target. With $p=30$ GeV, 
the momentum of a $\J$ produced at midrapidity at SPS energies
($E_{lab}=200$ AGeV), this yields $l_f\approx 3$ fm. 
Due to the lack of better knowledge, we use the same $l_f\approx 3$ fm for the 
$\chi$. For the $\Psi'$ we use $l_f\approx 6$ fm, i.e. we introduced an 
additional factor of 2, because it is not a small object, but has the size of 
a normal hadron, i.e. the pion. For $E_{lab}=800$ AGeV we get another factor 
of two for the formation lengths due to the larger Lorentz factor.

\begin{figure}
\centerline{\hbox{\psfig{figure=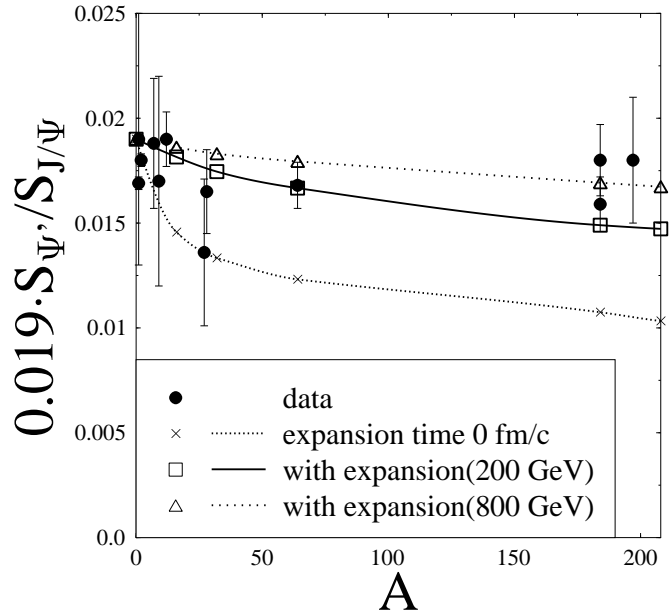,height=10cm}}}
\caption{The ratio $0.019\cdot S_{\Psi'} / S_{\J}$ is shown in $pA$ with
the calculated cross sections for the J/$\Psi$, $\Psi'$ and $\chi$ (crosses) 
in comparison to the data (circles). The squares and the triangles shows the
ratio calculated with the expansion of small wave packages
(see text).}
\label{ratio}
\end{figure}

However, this has a large impact on the 
$\Psi'$ to $\J$-ratio depicted in Fig.~\ref{ratio}, which
shows the ratio $0.019\cdot S_{\Psi'} / S_{\J}$ calculated with (squares 
(200 GeV) and triangles (800 GeV)) 
and without (crosses) expansion. The factor 0.019 is the measured value in 
$pp$ collisions, because the experiments do not measure the calculated value
$S_{\Psi'} / S_{\J}$ but 
$\frac{B_{\mu\mu}\sigma(\Psi')}{B_{\mu\mu}\sigma(\J)}$. 
$B_{\mu\mu}$ are the branching ratios for $\J ,\,\Psi'\rightarrow\mu\mu$. 
The calculations which take into
account the expansion of small wave packages show better agreement 
with the data (circles) (taken from~\cite{carlos}) than the calculation 
without expansion time, i.e. with immediate $\J$ formation, $l_f=0$.
We calculated this effect both at $E_{lab}=200$ AGeV and 800 AGeV. 
The data have been measured at different energies ($E_{lab}$ = 200, 300, 400, 
450, 800 GeV and $\sqrt{s}$ = 63 GeV). 

The charmonium survival probability $S^X_{AB}$ in minimum bias $AB$ collisions 
can be calculated by~\cite{hufner} $S^X_{AB}=S^X_A\cdot S^X_B$.
$S^X_A$ and $S^X_B$ are defined by eq.~\ref{glaub}. The neglect of the 
stopping of nucleons due to the energy loss seems safe: Drell-Yan processes 
do not suffer from this energy loss (see~\cite{kharzeev2,baier} 
and refs. therein).

The strong spin dependence of the $\chi$-$N$ cross section is due to the 
angular dependence of the wave functions, which leads to different transverse
quark separation for different states. The S-states (J/$\Psi$ and $\Psi'$)
in contrast, do not exhibit any dependence on the angles. 

\begin{figure}
\hspace*{-4cm} 
\centerline{\hbox{\psfig{figure=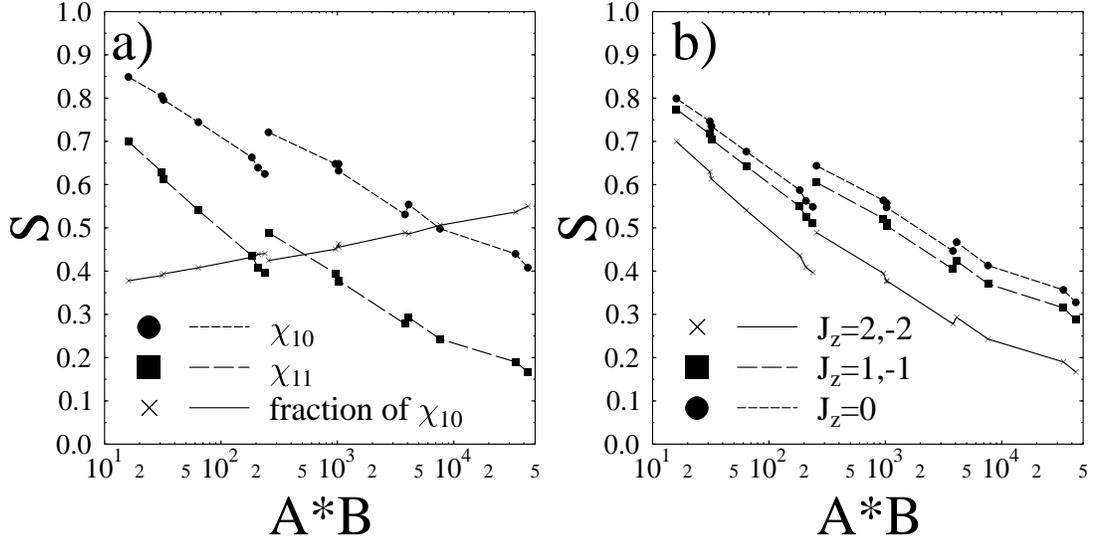,height=15cm,angle=-90}}}
\vspace*{-3cm}
\caption{The survival probabilities $S$ for the different $\chi$-states 
are shown for $pA$ and minimum bias $AB$ collisions. a) gives the 
$\chi_{10}$ and the $\chi_{11(-1)}$, which differ by 0.2-0.3. The 
fraction of the $\chi_{10}$'s of all $\chi$'s increases to more than 50 \% 
for $PbPb$-collisions. b) the $\chi$-states with J=2 and ${\rm J}_z=0,1(-1)$ 
and $2(-2)$ are shown.}
\label{chi}
\end{figure}

However, the P-states yield two vastly different 
cross sections (see Tab.~\ref{meanb}) for $\chi_{10}$ and $\chi_{11}$, 
respectively. This leads to a higher absorption rate of the $\chi_{11}$ as 
compared to the $\chi_{10}$. This new form of color filtering is predicted 
also for the corresponding states of other hadrons; e.g. for the bottomium 
states which are proposed as contrast signals to the $\J$'s at RHIC and LHC! 
(For a detailed review of color coherence effects 
see~\cite{lonya}.) Fig.~\ref{chi}a depicts the $A$ dependence of the
suppression of the $\chi_{11}$ and $\chi_{10}$ in 
$pA$ and in minimum bias $AB$ collisions.

The polarization of the $\chi$ shown in Fig.~\ref{chi}a may be 
difficult to observe. However, it is manifested in the production of all 
P-states (except in the $J^{PC}=0^{++}$ state, which in any case has a very 
small $\chi \rightarrow \J + \gamma$ branching ratio). Significant effects
are expected for both $1^{++}$ and $2^{++}$ states, which give the major 
contribution to the $\J$ production, and also  for the S=0 state $1^{+-}$.
This state was observed to date only in $p\bar p$ collisions~\cite{armstrong}. 
The predicted effect is present for the $1^{++}$ state, since the probability 
of the m=0 state is 1/2 for helicity +1 and is 0 for helicity 0.

There are other promising $\chi$ states with the total angular 
momentum J=2. 
Fig.~\ref{chi}b shows the different suppressions for these $\chi$-states, i.e.
with J=2 and J${}_z=0,1(-1)$ and $2(-2)$. These states are mixtures of 
$\chi_{10}$, $\chi_{11}$ and $\chi_{1-1}$: $\chi({\rm J,J}_z)=a_1\cdot 
\chi_{10} + a_2\cdot \chi_{11}+a_3\cdot \chi_{1-1}$. The $a$'s are given by the 
Clebsch-Gordan coefficients.

The measured $\J$-suppression up to sulphur-uranium
collisions seems to be described reasonably well with the present simple model,
in view of the neglect of comoving hadrons~\cite{neubauer,gavin}. 
Note that the physically most 
reasonable scenario, namely the one including the expansion 
actually underestimates the suppression for $SU$ by about 10 \%. 
And also the apparently strong $\J$ suppression found
at CERN in going from $SU$ to $PbPb$ collisions~\cite{Go96} can not be
explained in our calculations.
This leaves space for more physics.
Density fluctuations and the energy loss due to gluon radiation
have also been neglected.
Higher statistics for the $pA$-data are needed to clarify the importance
of the expansion effects. For example the ratio of $\Psi'$'s to $\J$'s should 
be measured at one energy
(e.g. $\sqrt{s}=20$ GeV, because here the Lorentz factor is not too large) and 
for different systems. This can prove that this ratio is not constant. 
One should also measure the rapidity distribution of this ratio to see the 
dependence on the Lorentz factor.

We recommend to measure the $A$-dependent polarization of the $\chi$ 
also in $pA$ reactions at $\sqrt{s}=20$ GeV
and in the midrapidity region, because the above described expansion effects
and the interaction with secondaries will decrease this polarization.
The polarization, once found, can be used to measure the interaction of these 
charmonium states with secondaries in $AB$ collisions.
The quantitative evaluation of the interaction with comovers, nucleon 
correlations and density fluctuations is postponed to the future.

The other new effect predicted by the QCD factorization theorem is the fast 
increase of the cross section
of spatially small configurations with energy, see Tab.~\ref{meanb}.
Thus, QCD predicts a large nuclear absorption of charmonium in $AB$ collisions 
at the RHIC and, in particular, at LHC energies, although the effective 
size of $c\bar c$ pairs will be $\sim \frac{1}{m_c}\sim 0.1$ fm
due to the strong increase of the parton distribution functions at small $x$. 
The absorption depends also  on the rapidity of the produced 
$c\bar c$ pair. 

\acknowledgements
We wish to express our sincere gratitude to Werner Koepf for providing 
his charmonium wave function program and to M. Gorenstein and C. Lourenco for 
discussions. This work was supported by the GSI, the 
Graduiertenkolleg "Theoretische und Experimentelle Schwerionenphysik", 
the DFG, the BMBF, the U.S. Department of Energy under Contract 
No.\ DE-FG02-93ER40771, the Alexander von Humboldt Foundation and 
the Israel Academy of Science under contract N 19-971.


\pagebreak
\begin{table}
\begin{center}
\begin{tabular}{|c|c|c|c|c|c|c|c|c|}
\hline
$c\overline{c}$/$b\overline{b}$-state & J/$\Psi$ & $\Psi'$ & $\chi_{c10}$ & $\chi_{c11}$ & $\Upsilon$ & $\Upsilon'$ & $\chi_{b10}$ & $\chi_{b11}$\\ 
\hline
$<b^2>$ (fm${}^2$) & 0.094 & 0.385 & 0.147 & 0.293 & 0.027 & 0.15 & 0.059 & 0.12 \\
\hline
$\sigma_{nonperturbative}$ (mb) & 3.62 & 20.0 & 6.82 & 15.9 & 0.31 & 7.35 & 1.48 & 4.48\\
\hline
$\sigma_{hard}$ (mb) (SPS)& 0.024 & 0.012 & 0.021 & 0.006 & 0.101 & 0.030 & 0.103 & 0.064\\
\hline
$\sigma_{hard}$ (mb) (RHIC)& 1.73 & 0.68 & 1.23 & 0.30 & 2.30 & 0.64 & 2.17 & 1.27\\
\hline$
\sigma_{hard}$ (mb) (LHC)& 20.8 & 8.2 & 14.7 & 3.5 & 28.4 & 7.8 & 26.3 & 15.0 \\
\hline
\end{tabular}  
\end{center}   
\vspace*{-.5cm}
\caption
{\label{meanb}The average square of the transverse distances of the charmonium 
states and the total quarkonium-nucleon cross sections $\sigma$.
For the $\chi$ two values arise, due to the spin dependent wave functions
($lm=10,11$). $\sigma_{hard}$ are the perturbative QCD contribution 
at different energies (see text).}
\end{table}

\end{document}